%% file: atotal.tex
\documentclass[12pt]{article}
\usepackage{graphicx}
\usepackage{caption}
\usepackage{subcaption}
\usepackage{amsmath,amssymb,mathrsfs,float,afterpage} 
\usepackage[longnamesfirst]{natbib}
\usepackage{tikz}

\begin{document}

\pagestyle{plain}
\parindent 0mm

\input{definitions}
\begin{center}
{\bf Optimal Designs for Spherical Harmonic Regression}\\
Linda M. Haines\\
{\small Department of Statistical Sciences,\\
University of Cape Town, Rondebosch 7700, South Africa.\\
email: linda.haines@uct.ac.za ~~~ Orchid: 0000-0002-8843-5353}
\end{center}

\noindent{\bf Abstract}

This short paper is concerned with the use of spherical t-designs as optimal designs for the spherical harmonic regression model in three dimensions over a range of specified criteria. The nature of the designs is explored  and their availability and suitability is reviewed

\bigskip

\noindent{\bf 1. Introduction}

Spherical harmonic regression models play a key role in a broad spectrum of fields, ranging from cell biology to ambisonics and geophysics. At the same time, optimal designs for the model have received scant attention in the statistical literature, as evidenced by  the two early papers of \citet{kup:73} and \citet{lc:75} and the more recent papers of \citet{dmp:05}, \citet{dw:09} and \citet{dksg:19}. The results introduced in these papers hinge on the fact that optimal designs for the spherical harmonic regression model over a wide range of criteria are given by the uniform distribution over the sphere. Such designs cannot be implemented in practice and  optimal exact designs which put equal weights on the support points were necessarily constructed.  \citet{dksg:19}, in their paper on hyperspherical harmonic regression, noted that spherical $t$-designs constitute exact counterparts to the uniform distribution on the sphere and could therefore be used as optimal designs. However, the authors  expressed concern that identifying spherical $t$-designs for use with a fixed number of support points may not be possible within the spherical harmonic context and did not explore the issue  further.

The present study focusses on spherical harmonic regression in three dimensions, that is on the sphere $S^2$, and reviews the wealth of spherical $t$-designs readily available for use as optimal exact designs in the modelling process. The paper is structured as follows. Section 2 contains a brief discussion of the spherical harmonic model and the relevant optimal design literature. Spherical $t$-designs are then introduced formally in Section 3 and their construction reviewed. An outline of the stereographic projection used to highlight the structure of the designs of interest is also included in that section. Illustrative examples are provided in Section 4 and focus primarily on spherical harmonic expansions up to order $7$. A brief discussion of the findings of the study is given in Section 5. 
\bigskip
\bigskip

\noindent{\bf 2. Preliminaries}

{\em 2.1 Spherical Harmonic Regresson}

To fix ideas, suppose that the distances $y_i$ from a sensor at the origin to the boundary of a 3D shape of interest are measured in directions specified by the polar angle $\theta_i$ and the azimuthal angle $\phi_i$, where $\theta_i \in [0, \pi]$ and $\phi_i \in (-\pi,\pi]$, $i=1 \ldots, n$. Spherical harmonic descriptors can then be used to fit the resultant data by invoking the  regression model 
\begin{equation}
y_i = \sum_{\ell=0}^d \sum_{m=-\ell}^{\ell} c_{\ell}^m Y_{\ell}^{m} (\theta_i, \phi_i) + e_i, i=1, \ldots n,
\label{model}
\end{equation}
where the terms $Y_{\ell}^{m} (\theta_i, \phi_i)$ are real spherical harmonics, the  terms $c_{\ell}^m$ comprise $(d+1)^2$ unknown regression coefficients and $e_i$ is an error term distributed as $N(0,1)$. More specifically, the real spherical harmonics can be expressed as
\begin{eqnarray*}
Y_{\ell}^{0}(\theta, \phi) &=& \sqrt{2 \ell +1}\; P_{\ell}^{0}(\cos\theta), \; \ell \in \mathbb{N}_0; m=0,\\ 
Y_{\ell}^{m}(\theta, \phi) &=& \sqrt{2 (2 \ell +1) \frac{(\ell-|m|)!}{(\ell+|m|)!}} \; P_{\ell}^{|m|}(\cos\theta) \sin(m \phi), \; \ell> 0, \; m=-\ell, \ldots, -1,\\
Y_{\ell}^{m}(\theta, \phi) &=& \sqrt{2 (2 \ell +1) \frac{(\ell-m)!}{(\ell+m)!}} \; P_{\ell}^{m}(\cos\theta)  \cos(m \phi), \; \ell> 0, \; m=1, \ldots, \ell,
\end{eqnarray*}
where $P_{\ell}^m(x)$ denotes the $m$th associated Legendre polynomial of degree $\ell$.  The index $d$ in the outer summation of model (\ref{model}) represents the order of the truncated expansion imposed on the underlying infinite sum and hence the order of the model.  In addition, the angles $\theta$ and $\phi$ define a direction in $\mathbb{R}^3$ from the centre of the  sphere  and are specified here, and in general, as unique points on the unit  sphere $S^2$ itself.
\bigskip

{\em 2.2 Optimal Designs}

The question now arises as to how the polar and azimuthal angles, $\theta_i$ and $\phi_i$, $i=1, \ldots, n,$ should best be chosen in order to obtain precise estimates of the $(d+1)^2$ unknown parameters of the model, $c_0^0, c_{1}^{-1}, c_1^{0}, c_1^1, \ldots,  c_d^{d}$. 
\citet{dmp:05} and \citet{dw:09} addressed this problem by constructing designs which were, in some sense, optimal with respect to the precision of the estimates of the unknown parameters of the model. Specifically, the authors considered  Kiefer's suite of $\Phi_p$-criteria and an integrated mean square error criterion and demonstrated that the optimal designs so constructed placed a uniform distribution $\xi^*=\displaystyle \frac{1}{4 \pi} \sin\theta \: d\theta \: d \phi$ on the surface of the sphere $S^2$, with  information matrix equal to the identity $I_{(d+1)^2}$. 
Clearly, designs with distribution $\xi^*$ on the sphere cannot be implemented in practice and \citet{dmp:05} therefore sought both exact and approximate  optimal designs given by a product measure  $\mu \otimes \nu$ for which the  information matrix $M(\mu \otimes \nu)$ is the identity.  
In the exact case, which is of interest here, the authors  took the measure $\mu$ to comprise polar angles as $n_{\theta}$ equally-weighted support points  drawn from the literature on univariate quadrature for $d=1, 2,3$ and $4$ and calculated numerically for $d = 5, 6$ and $7$, and the measure $\nu$ to comprise  $n_{\phi}$ equally-spaced points specified by the azimuthal angles $\phi_j = \alpha + \displaystyle \frac{2 \pi j}{n_{\phi}}$, where $\alpha \in (-\pi, \pi], \; j=1, \ldots, n_{\phi},$ and $n_{\phi} \ge 2 d+1$. The results are reported in Table 1 of their paper and indicate that, because the lower limit placed on $n_{\phi}$ is high, the total numbers of points of the designs increase rapidly with increasing order $d$ so as to become unworkable. 
\bigskip
\bigskip

{\bf 3. Spherical $t$-designs}

{\em 3.1 Key Result}

Spherical $t$-designs were developed by Delsarte, Goethals and Seidel in a seminal paper in 1977 and have been, and continue to be, extensively researched.  In the present context of three dimensions, a spherical $t$-design is a set of points on the unit sphere $S^2$ such that the average of the points over a homogeneous spherical polynomial of specified degree is exactly equal to the integral of that polynomial with respect  to a uniform measure over the sphere. More formally, a spherical $t$-design is a finite subset $Y$ on the unit sphere $S^{2} \subset \mathbb{R}^{3}$ which satisfies the cubature relationship 
\begin{equation}
\int_{S^{2}}   f(x) d\sigma(x) = \frac{1}{|Y|} \sum_{y \in Y} f(y) ~~ \mbox{for~all~} f(x) \in \mbox{Pol}_t(S^{2}),
\label{spht}
\end{equation}
where $\sigma$ is a uniform measure on the sphere and $\mbox{Pol}_t(S^{2})$ denotes the space of homogeneous spherical polynomials of degree less than or equal to $t$.  Lower bounds on the number of points in such a design are well-documented and, for the sphere $S^2$, are given by $
\frac{(t+2)^2}{4} \mbox{ if $t$ is even and } \frac{(t+1) (t+3)}{4} \mbox{ if $t$ is odd} 
$ \citep{hsw:10}.
These bounds are only attained for designs with $t=1,2,3$ and $5$. However, it has been shown that  spherical $t$-designs always exist provided the number of points is sufficiently large. 
The key result  relating to this paper now follows immediately.
\medskip

\noindent{\bf Result:} Spherical $t$-designs on the sphere $S^2$ with $t \ge 2 d$ are optimal for a spherical harmonic regression model of order $d$ with respect to Kiefer's suite of $\Phi_p$-optimality  criteria and to other criteria specified by \citet{dmp:05} and  \citet{dw:09}.
\medskip

The exact designs given in Table 1 of  \citet{dmp:05} are therefore spherical $t$-designs. Specifically, the designs  were, in effect, constructed by invoking the longitude-latitude rule for numerical integration on the sphere $S^2$ \citep{hsw:10}, an approach which is closely related  to that used by \citet{baj:91} within the context of algebraic combinatorics. 
As an aside, \citet{dmp:05} constructed optimal approximate designs for the spherical harmonic regression in three dimensions, that is designs which place weights which sum to one but cannot necessarily be used as the basis for  exact designs with a specified total number of points. In fact, \citet{gp:11}  present designs based on ``Gauss-type'' cubature over $S^2$  which can be regarded as interesting alternatives to the optimal approximate designs of \citet{dmp:05} and which  are available on the website \citet{graefweb}.
\bigskip
\newpage
{\em 3.2 Designs}

Certain spherical $t$-designs are immediately available as, for example, those with points taken to be the vertices of a Platonic solid inscribed in the unit sphere, $S^2$. Others have been obtained explicitly and are reported in the  literature on algebraic combinatorics. However such examples are not common  and spherical $t$-designs must in general be obtained numerically.  In fact, the construction of spherical $t$-designs over the unit sphere is  fraught with computational problems and, more specifically, as stated in \citet{wom:18},  ``$\ldots$ optimization problems with points on the sphere typically have many different local minima with different characteristics.'' As a consequence only a limited number of  numerically-based studies, those of  \citet{hs:96}, \citet{gp:11} and \citet{wom:18},  have been reported.
 \bigskip
 
\citet{hs:96}, in Table 1 of their paper, provide a catalogue of spherical $t$-designs for which, for a given  number of support points ranging from 1 to 100, the value of $t$ is a maximum and, thereby, covers values of $t$ from 1 to 13. 
 Thus, spherical $t$-designs for a given $t$ can be drawn from the catalogue and include, by the nature of the construction, the design with the smallest number of points. The table also includes the point groups to which the specified designs belong, together with the  orbits of the support points, and, in certain cases, the names of the  attendant convex polyhedra. 
The spherical $t$-designs from the catalogue of Hardin and Sloane, together with designs for $t >13$, two  for $t=15$ and single designs for $t=14, 16,  \ldots, 21$, can be obtained from  Neil Sloane's website, \citet{nsweb}, with points recorded up to 16 decimal places. 
\bigskip

The spherical $t$-designs of \citet{hs:96} emerged from a study into the construction of  $I$-optimal designs for quadratic polynomial regression. In contrast, \citet{gp:11} and \citet{wom:18} considered spherical $t$-designs for use in numerical integration on the sphere  $S^2$. As a consequence, the latter authors constructed  just one  design for a given value of $t$ but considered  a vast range of such $t$ values. In particular, \citet{gp:11} constructed designs which have the minimum numbers of points for selected values of $t$, identified the attendant point groups and displayed the designs in three dimensions on the website \citet{graefweb}.  \citet{wom:18}  used a bespoke optimization routine to build spherical $t$-designs designs with $t=1, \ldots, 180$ and antipodal designs with $t=1,3, \ldots, 325$ which are optimal with respect to specific geometric properties on the sphere. These designs, with points given to a very high accuracy, are all immediately available on the websites \citet{graefweb} and \citet{womweb}.
\bigskip

{\em 3.3 Stereographic Projection}

The stereographic projection is invoked in the present study in order to showcase the symmetry inherent in selected spherical $t$-designs and, thereby, to make the designs more attractive to the practitioner. Specifically, the projection  maps points on the sphere in three dimensions, taken to be a  perfect terrestrial globe, onto the two-dimensional equatorial plane and the resultant construct is termed a stereogram. The process is described in detail in books on crystallography such as that of \citet{hammond:15} and on design in the paper by \citet{haines:24}.
Thus, a point on the sphere in the northern hemisphere is represented on the stereogram  by the point of intersection of the line joining that point to the south pole  and is indicated there by a solid circle. Similarly, a point on the sphere in the southern hemisphere is represented on the stereogram  by the  point of intersection of the line joining the point to the north pole and is represented by an open circle. Meridians, that is longitudes, are represented on the stereogram by lines through the origin and parallels, that is latitudes, by circles centred at the origin. In addition, great circles map onto the stereogram as arcs with end-points diametrically opposite and small circles map onto circles but the centres of the two circles do not necessarily correspond. Directions from the origin of the unit sphere in three-dimensions are indicated on the stereogram by using crystallographic notation and, for example,  the vector $(-1,1,0)$ is represented compactly as $[\bar{1}10]$.
\bigskip
\bigskip

{\bf 4. Examples}

Details of selected spherical $t$-designs for the spherical harmonic regression model (1) with order of truncation $d=1, \ldots, 7$, number of parameters $(d+1)^2$ and the smallest spherical $t$-design permitted for a given $d$, that is $t_{\min}=2d$, are presented in Table 1. The designs recorded  in the table under `Hardin + Sloane' are taken from  \cite{hs:96} with the minimum number of points for $t=2d$ and $t=2 d+1$ denoted $n_{2 d}$ and $n_{2d+1}$. 
The designs under the heading `Dette et al'  are from the paper by \citet{dmp:05} with  the numbers of polar and azimuthal angles given by $n_{\theta}$ and $n_{\phi}$, where $n_{\phi} $ is  taken to be $2 d+1$ so that the total number of design points,  $n_{tot}$,  is a minimum.
\begin{table}[ht]
\centering
\begin{tabular}{|c|c|c|cc|c|cc|c|} \hline
\multicolumn{3}{|c|}{Parameters} &    \multicolumn{3}{c|}{Hardin + Sloane} & \multicolumn{3}{c|}{Dette et al}\\
$d$ & $(d+1)^2$ & $t_{\min}$  & $n_{2d}$ & $n_{2d+1}$ & $n_{min}$ & $n_{\theta}$ & $n_{\phi}$ &  $n_{tot}$\\ \hline
$1$ & $4$ & $2$ & $4$ & $6$ & $4$ & 2 & 3 & $6$\\ \hline
$2$ & $9$ & $4$ & $14$ & $12$ & $12$ & $4$ & $5$ & $20$\\ \hline
$3$ & $16$ & $6$   & $26$ & $24$ & $24$ & $6$ & $7$ & $42$\\ \hline
$4$ & $25$ & $8$  & $36$ & $48$ & $36$ & $9$ & $9$ & $81$\\  \hline 
$5$ & $36$ & $10$  & $60$ & $70$ & $60$ & $13$ & $11$ & $143$ \\ \hline
$6$ & $49$ & $12$   & $84$ & $94$ & $84$ & $17$ & $13$ & $221$\\ \hline
$7$ & $64$ & $14$   & $108$ & $120$ & $108$ & $23$ & $15$ & $345$ \\ \hline
\end{tabular}
\caption{Summary of the settings for $d=1, \ldots, 7$ with minimum numbers of points}
\end{table} 
\bigskip

\begin{figure}[ht] 
\vspace{-50mm}

\hspace{-40mm} \includegraphics{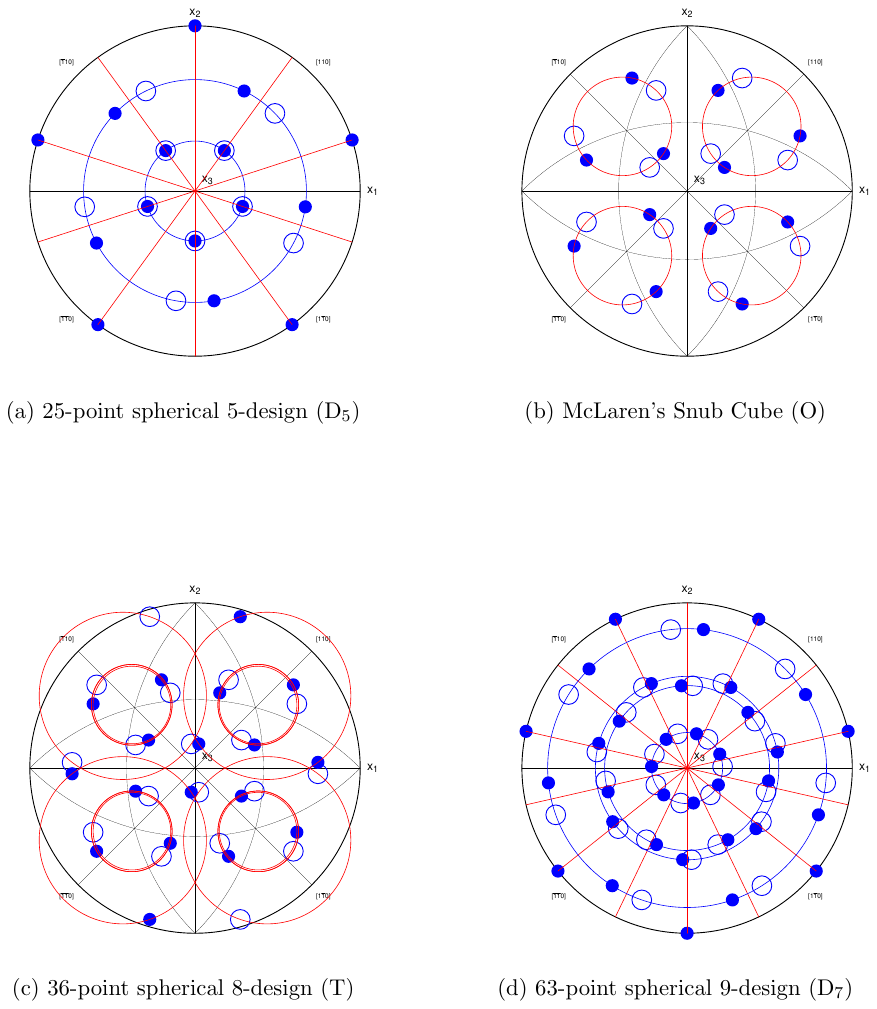} 
\vspace{-75mm} 

\caption{Stereograms of selected designs with point groups in the Sch\"{o}nflies notation in brackets. McLaren's Snub Cube is a 24-point spherical 7-design.}\label{fig1}
 \end{figure}

The spherical $t$-designs of \cite{hs:96}, and also those on the website \citet{graefweb}, have interesting symmetries. Specifically, the designs with minimal points for $d=1$ and $d=2$ correspond to the vertices of the Platonic solids inscribed in the sphere $S^2$, that is the tetrahedron and the octahedron for $d=1$ and the icosahedron and dodecahedron for $d=2$. A large number of spherical 4- and 5-designs is in fact  available. To illustrate, the stereogram of a 25-point spherical 5-design is shown in Figure \ref{fig1}(a) and highlights the inherent five-fold symmetry of the design. For $d=3$, the $24$-point spherical $7$-design which \citet{hs:96} referred to as ``McLaren's improved snub cube'' is particularly attractive.  The authors record moving the vertices of the regular snub cube inscribed in the sphere ``slightly'' over the sphere and demonstrated, by a subtle use of algebra, that the resultant design is a spherical $7$-design with points determined by a cubic polynomial. \citet{hs:96} noted however that the design itself had been reported by McLaren in 1963 \nocite{mcl:63} using arguments based on point groups. A stereogram of this spherical $7$-design is given in Figure \ref{fig1}(b) and illustrates the rotation of the points around the three-fold axes. For $d=4$, spherical $t$-designs for $t=8$ and $t=9$ with the minimum numbers of points  were constructed by \citet{hs:96} algebraically. Specifically, the spherical $8$-design has $36$ points on the sphere arranged as three snub tetrahedra and the spherical $9$-design has $48$ points defined by two snub cubes. A stereogram of the $36$-point spherical $8$-design is shown in Figure \ref{fig1}(c) and illustrates clearly the nature of designs which are specified by snub tetrahedra.
The optimal spherical $t$-designs with a minimum number of points for  $d=5, 6$ and $7$, that is with $t=10, 12$ and $14$, correspond  to $5$, $7$ and $9$ snub tetrahedra, respectively, but are defined by  large numbers of polar angles. 
\bigskip

For $d > 7$, the number of parameters of the spherical harmonic regression model rapidly becomes excessive and, for example, for $d\ge 10$ the number exceeds $100$.  Spherical $t$-designs for such cases are available from the websites \citet{graefweb} and \citet{womweb} but, as noted in Subsection 3.2,  are restricted to one design for each value of $t$.  Specifically,  \citet{graefweb}  presents designs for which  the  number of points for each value of $t$ is a minimum and the designs of \citet{womweb} coincide or differ from those designs by a single point. A crucial feature of the designs however is that the total numbers of points are less than, but very close to, twice the number of parameters. In the context of degrees of freedom for the regression model, it is therefore, arguably, not worthwhile searching for  spherical $t$-designs with more points than the minimum and  the designs of \citet{graefweb} can reasonably be used. 
It also follows from the Result presented in Section 3.1  that any  spherical $t$-design with $t$ strictly greater than $2 d$ is itself a spherical $2 d$-design and can therefore be used  in the present context as an optimal design for the spherical harmonic regression model of order $d$ .
\bigskip
\bigskip

{\bf 5. Discussion}

This paper examines the use of spherical $t$-designs with $t \ge 2 d$ as optimal exact designs in three dimensions, that is on the sphere $S^2$, for the spherical harmonic regression model of order $d$. Specifically, for a spherical harmonic expansion of order less than or equal to $7$  the catalogue of designs introduced by \citet{hs:96}  and available on Neil Sloane's website \citep{nsweb} is demonstrated to be  comprehensive and invaluable. In the present study, the catalogue is reinforced by invoking the stereographic projection and attendant stereogram in order to  provide a neat schematic in two dimensions from which a researcher can visualise the design structure. 
In addition, the spherical $t$-designs of \citet{hs:96} are to be preferred to those of \citet{dmp:05} in that the former comprise fewer total numbers of points and can be selected to accommodate a reasonably small number of polar angles. 
For orders of expansion of the spherical harmonic regression model greater than  $7$,  there would seem to be only two sources of spherical $t$-designs from which to draw, those of \citet{gp:11} with website \citet{graefweb} and of \citet{wom:18} with website \citet{womweb} and, together, these provide only  two or three designs with a fixed number of points for a given value of $t$. However the minimum number of points for a spherical $t$-design in this range, as given in \citet{graefweb}, was found to be very  close to twice the number of parameters in the model and, in the context of degrees of freedom, the  minimum point design used in the modelling process  should therefore suffice. In any case, it should be stressed that any spherical $t$-design for which $t$ is greater than $2 d$ can also be used for models with order of truncation $d$.
\bigskip

\cite{dksg:19} extended the work of \citet{dmp:05} to higher dimensions, that is to optimal designs for hyperspherical harmonic regression in dimensions greater than three.   
The authors identified the optimal exact designs required in practice as hyperspherical $t$-designs but highlighted problems relating to the availability of such designs.   
In fact, for hyperspherical harmonic regression on the  sphere $S^3$, that is in  four dimensions,  the requisite spherical $t$-designs are discussed in some depth by \citet{wom:18} and  are immediately available on the website \citet{womweb}. The designs  include the six regular convex polytopes in $\mathbb{R}^3$, as, for example, the well-known 600-cell and 120-cell designs with $t=11$, and a range of designs which are either non-antipodal or antipodal. If, as in the case of the present study on the sphere $S^2$, these spherical $t$-designs have a minimum numbers of points which are close to twice the number of parameters of the associated model, they could well suffice. Alternatively, designs with $t$ greater than $2 d$ could be invoked.
\bigskip
\bigskip

{\bf Acknowledgements}

I would like to thank two anonymous referees for their helpful comments.
I would also like to thank the University of Cape Town and the National Research Foundation (NRF) of South Africa, grant (UID) 119122, for financial support. Any opinion, finding and conclusion or recommendation expressed in this material is that of the author and the NRF does not accept liability in this regard.

\bibliographystyle{chicago}	   
\bibliography{spherical.bib}

\begin{figure}[ht] 
\vspace{-10mm}

\hspace{-40mm} \includegraphics{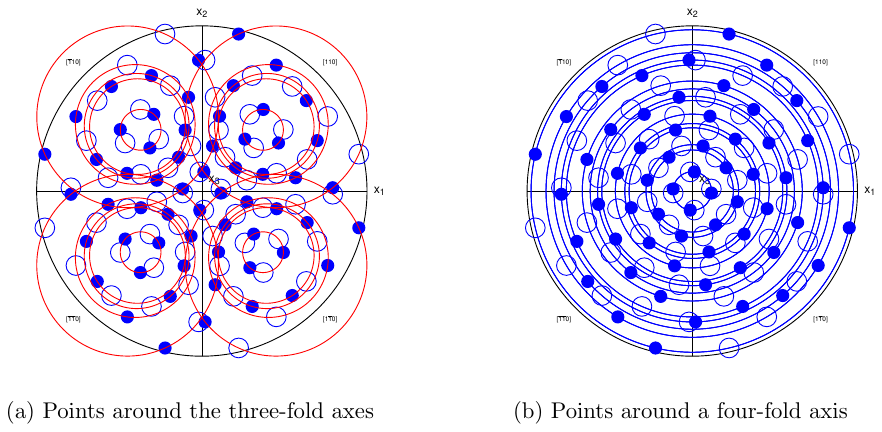} 
\vspace{-180mm} 

\caption{Stereograms of the 120-point spherical 15-design with point group O in Sch\"{o}nflies notation }\label{fig2} \end{figure}
\end{document}

%% file: definitions.tex
\newcommand{\spc}[1]{\Pisymbol{cryst}{#1}}
\newcommand{\spctf}[1]{\begin{turn}{45} \Pisymbol{cryst}{#1}
\end{turn}} 
\newcommand{\spctn}[1]{\begin{turn}{90} \Pisymbol{cryst}{#1}
\end{turn}}

\def\bA{\mbox{\boldmath$A$}}
\def\ba{\mbox{\boldmath$a$}}
\def\bB{\mbox{\boldmath$B$}}
\def\bb{\mbox{\boldmath$b$}}
\def\bC{\mbox{\boldmath$C$}}
\def\bc{\mbox{\boldmath$c$}}
\def\bD{\mbox{\boldmath$D$}}
\def\bd{\mbox{\boldmath$d$}}
\def\bE{\mbox{\boldmath$E$}}
\def\be{\mbox{\boldmath$e$}}
\def\bF{\mbox{\boldmath$F$}}
\def\boldf{\mbox{\boldmath$f$}}
\def\bG{\mbox{\boldmath$G$}}
\def\bH{\mbox{\boldmath$H$}}
\def\bh{\mbox{\boldmath$h$}}
\def\bI{\mbox{\boldmath$I$}}
\def\bJ{\mbox{\boldmath$J$}}
\def\bK{\mbox{\boldmath$K$}}
\def\bk{\mbox{\boldmath$k$}}
\def\bL{\mbox{\boldmath$L$}}
\def\bM{\mbox{\boldmath$M$}}
\def\bN{\mbox{\boldmath$N$}}
\def\bn{\mbox{\boldmath$n$}}
\def\bP{\mbox{\boldmath$P$}}
\def\bp{\mbox{\boldmath$p$}}
\def\bq{\mbox{\boldmath$q$}}
\def\bR{\mbox{\boldmath$R$}}
\def\br{\mbox{\boldmath$r$}}
\def\bS{\mbox{\boldmath$S$}}
\def\bT{\mbox{\boldmath$T$}}
\def\bt{\mbox{\boldmath$t$}}
\def\bU{\mbox{\boldmath$U$}}
\def\bu{\mbox{\boldmath$u$}}
\def\bv{\mbox{\boldmath$v$}}
\def\bW{\mbox{\boldmath$W$}}
\def\bw{\mbox{\boldmath$w$}}
\def\bX{\mbox{\boldmath$X$}}
\def\bx{\mbox{\boldmath$x$}}
\def\by{\mbox{\boldmath$y$}}
\def\bZ{\mbox{\boldmath$Z$}}
\def\bz{\mbox{\boldmath$z$}}

\def\bbeta{\mbox{\boldmath$\beta$}}
\def\beps{\mbox{\boldmath$\epsilon$}}
\def\bSigma{\mbox{\boldmath $\Sigma$}}
\def\btau{\mbox{\boldmath $\tau$}}
\def\btheta{\mbox{\boldmath $\theta$}}
\def\blam{\mbox{\boldmath $\lambda$}}

\def\cc{{\cal C}}
\def\calr{{\cal R}}
\def\cw{{\cal W}}
\def\cx{{\cal X}}
\def\cz{{\cal Z}}

\def\bzero{\mbox{\boldmath $0$}}
\def\b1{\mbox{\boldmath $1$}}
\def\bsim{\mbox{\boldmath $\sim$}}

\def \ni{\noindent}
\def \ds{\displaystyle}
\def \ul{\underline}
\def \fns{\footnotesize}
\def \ds{\displaystyle}

\newcommand{\hg}[2]{\mbox{}_{\scriptscriptstyle #1} F_{\scriptscriptstyle #2}}
\def \hlam{\hat{\lam}}
\def \hp{\hat{p}}
\newcommand{\overbar}[1]{\mkern 1.5mu\overline{\mkern-1.5mu#1\mkern-1.5mu}\mkern 1.5mu}

\def \mur{\frac{\mu}{\mu+r}}
\def \rmu{\frac{r}{\mu+r}}

\def \a{\alpha}
\def \b{\beta}
\def \s{\sigma}
\def \e{\epsilon}
\def \ul{\underline}
\def \t{\theta}
\def \ds{\displaystyle}
\def \d{\delta}
\def \g{\gamma}
\def \lam{\lambda}
\def \om{\omega}